\documentclass[journal]{IEEEtran}

\usepackage{amsfonts}
\usepackage{slashbox}
\usepackage{amsmath}
\usepackage{multirow}
\usepackage{graphicx}
\usepackage{amsfonts}
\usepackage{amsmath,epsfig}
\usepackage{mathbbold}
\usepackage{amssymb}
\usepackage{url}
\usepackage{bm}
\usepackage{cite}
\usepackage{amsmath}
\usepackage{amssymb}
\usepackage{latexsym}
\usepackage{multirow}
\usepackage{epsfig}
\usepackage{graphics}
\usepackage{mathrsfs}
\usepackage{algorithmic}
\usepackage{algorithm}
\usepackage{subfigure}
\usepackage{slashbox}
\usepackage[all]{xy}
\usepackage{setspace}

\usepackage{pifont}
\usepackage{bbding}
\usepackage{stmaryrd}
\usepackage{amssymb}
\usepackage{amsfonts}
\usepackage{epic}
\usepackage{latexsym}
\usepackage{epstopdf}
\usepackage{epic}
\usepackage{bm}
\usepackage{xcolor}
\usepackage{mathrsfs}
\usepackage{pifont}
\usepackage{bbding}
\usepackage{amsmath,epsfig}
\usepackage{mathbbold}
\usepackage{stmaryrd}
\usepackage{amssymb}
\usepackage{amsfonts}
\usepackage{epic}
\usepackage{graphicx}
\usepackage{subfigure}

\usepackage{enumerate}

\usepackage{graphicx}
\usepackage{subfig}
\usepackage{caption}

\usepackage{stfloats}
\usepackage{latexsym}
\usepackage{epstopdf}
\usepackage{epic}
\usepackage{multirow}

\usepackage{bm}

\usepackage{booktabs}
\usepackage{color}
\usepackage{cases}
\usepackage{array}
\usepackage{makecell}
\usepackage{times}

\graphicspath{{./figs/}}

\begin{document}
\title{Towards Smart and Reconfigurable Environment:  Intelligent Reflecting Surface Aided Wireless Network}
\author{\IEEEauthorblockN{Qingqing Wu and Rui Zhang } }

%
\maketitle
%

\begin{abstract}
Intelligent reflecting surface (IRS) is a new and revolutionizing technology that is able  to significantly improve the performance of wireless communication networks, by smartly reconfiguring the  wireless propagation environment with the use of massive  low-cost passive reflecting  elements integrated on a planar surface.  Specifically, different elements of an IRS can independently reflect the incident signal by controlling its amplitude and/or phase and thereby  collaboratively achieve fine-grained three-dimensional (3D) passive  beamforming for directional  signal enhancement or nulling.
  In this article, we provide an overview of the IRS technology, including  its main applications in wireless communication, competitive advantages over existing technologies,  hardware architecture as well as  the corresponding new signal model.  We focus on the key challenges in designing and implementing the new IRS-aided hybrid (with both active and passive components) wireless network, as compared to the traditional network comprising  active components only. Furthermore, numerical results are provided to show the great  performance enhancement with the use of IRS in typical wireless networks.
\end{abstract}

\vspace{-0.3cm}
\section{Introduction}

The targeted  1000-fold network capacity increase and ubiquitous wireless connectivity for at least 100 billion devices by  the forthcoming fifth-generation (5G) wireless  network have been largely achieved, thanks to the various key enabling technologies such as  ultra-dense network (UDN), massive multiple-input multiple-output (MIMO),  millimeter wave (mmWave) communication, and so on   \cite{boccardi2014five}. However, the required high complexity and  hardware cost as well as increased energy consumption are still crucial issues that remain unsolved. For instance, densely deploying base stations (BSs) or access points (APs) in a UDN  not only entails  increased hardware expenditure and maintenance cost, but also aggravates the network interference issue. In addition,  how to provide  reliable  and  scalable backhauls for UDN is a  challenging  task in practice, especially  for  indoor deployment  without the full optical  coverage  yet. Furthermore,  extending  massive MIMO from sub-6 GHz   to mmWave frequency bands generally  requires more complex signal processing as well as more costly  and energy consuming  hardware (e.g., radio frequency (RF) chains). Therefore, research on finding innovative, spectral and energy efficient, and yet
cost-effective solutions for future/beyond-5G  wireless networks is still imperative \cite{wu2016overview}.

In addition,  although  the 5G physical layer technologies are generally  capable of adapting to the space and time varying wireless environment, the signal propagation  is essentially random and  largely uncontrollable. Motivated by the above, intelligent reflecting surface (IRS) has been recently  proposed as a promising new  technology  for reconfiguring the wireless propagation environment via software-controlled reflection  \cite{wu2018IRS,JR:wu2018IRS,huangachievable,tan2018enabling}. {Specifically, IRS is a planar surface comprising   a large number of  low-cost passive reflecting  elements,  each being able to induce an amplitude and/or  phase change to  the incident signal independently,  thereby collaboratively achieving fine-grained three-dimensional (3D) reflect beamforming.
 In a sharp contrast to the existing wireless  link adaptation techniques at transmitter/receiver, IRS proactively modifies the wireless channel between them via highly controllable and intelligent signal reflection. This thus provides a new degree of freedom (DoF) to further enhance the wireless  link performance  and paves the way to realizing a smart and programmable wireless environment.
By properly adjusting the 3D passive beamforming,   the signal reflected by IRS can add constructively  with those  from the other paths  to enhance the desired signal power at the receiver,  or destructively to cancel the undesired signal such as  co-channel interference.}  {Since IRS eliminates the use of transmit RF chains and operates only in short range, it can be densely deployed with scalable  cost and low energy consumption, yet without the need of sophisticated interference management among passive IRSs.}

\begin{figure*}[!t]
\centering
\includegraphics[width=0.99\textwidth]{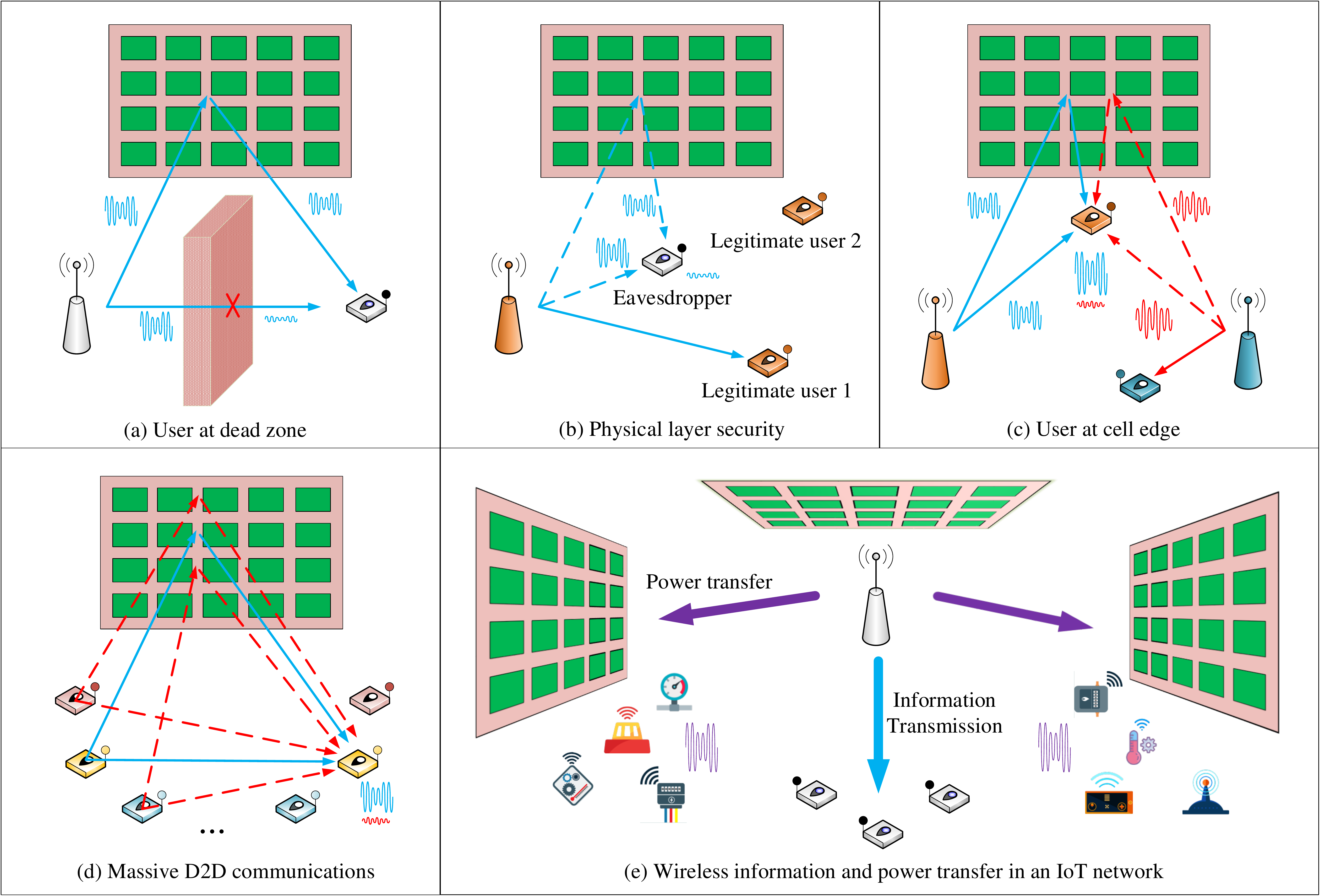}
\caption{Typical IRS applications in  wireless network.  } \vspace{-0.6cm}\label{SecI:APP}
\end{figure*}

Fig. \ref{SecI:APP} illustrates several typical  applications of the IRS-aided wireless network. In Fig. \ref{SecI:APP} (a),  a user is located in a dead zone where the direct link between it and its serving BS is severely blocked by an obstacle. In this case,  deploying an IRS that has clear links with the BS and user  helps  bypass the obstacle via intelligent signal  reflection and thus creates a virtual line-of-sight (LoS) link between them. This is particularly useful for the coverage extension in mmWave communications that are highly vulnerable to indoor blockage.  Fig. \ref{SecI:APP} (b) shows the use of  IRS for improving the physical layer security. When the link distance from the  BS to the eavesdropper is smaller than that to the legitimate user (e.g., user 1), or  the eavesdropper lies in the same direction as the legitimate user (e.g., user 2), the achievable secrecy communication rates are highly limited (even by employing transmit beamforming at the BS in the latter case). However, if an IRS is deployed in the vicinity of the eavesdropper,  the reflected signal by IRS can be tuned to cancel out  the (non-IRS-reflected)  signal from the BS at the eavesdropper, thus effectively reducing the information leakage.  In Fig. \ref{SecI:APP} (c), for a cell-edge user that suffers both high signal attenuation from its serving BS and severe co-channel interference from a neighboring BS, an IRS can be deployed at the cell edge to  help not only improve the desired signal power but also suppress the  interference by properly designing its reflect  beamforming,  thus  creating  a  ``signal  hotspot''  as  well  as ``interference-free zone''  in its vicinity. Fig. \ref{SecI:APP} (d) illustrates the use of  IRS for enabling massive device-to-device (D2D) communications where the IRS acts as a signal reflection hub to support simultaneous low-power transmissions via interference mitigation. Last, Fig. \ref{SecI:APP} (e) shows the application of  IRS for realizing  simultaneous wireless information and power transfer (SWIPT) to miscellaneous devices in an Internet-of-things (IoT) network \cite{bi2014wireless},  where the large aperture of IRS is leveraged to compensate  the significant power loss over long distance via passive  beamforming to nearby IoT devices to improve the efficiency of  wireless power transfer to them. Recently, there have been some works studying the above proposed applications \cite{JR:wu2019discreteIRS,wu2019weighted,guan2019intelligent}.

Besides the above   promising applications, IRS also possesses appealing  advantages  from an implementation viewpoint.  First, IRSs are usually  fabricated with low  profile,  light weight,  and  conformal geometry, which makes it easy to mount/remove them on/from the wall, ceiling, building facades, advertisement  panels,  etc. {Furthermore, since IRS is a complementary device in wireless networks, deploying it in existing  wireless systems  (e.g. cellular or WiFi) does not  require to  change their standardization and  hardware, while only necessary modification of the communication protocols suffices.}  As a result, the integration of IRS into wireless  networks can be made transparent to the users, thus providing high flexibility and superior compatibility with existing wireless systems. Therefore,  IRS can be practically deployed and integrated in wireless networks with low cost.

Next, we highlight the main differences as well as  competitive advantages of IRS as compared to  other existing  technologies related to IRS, namely,  active relay, backscatter communication, and active surface based massive MIMO \cite{hu2017beyond}. 
First,  compared to active wireless relay that assists in source-destination communication by signal regeneration and retransmission, IRS does not use any active  transmit module (e.g., power amplifier) but only reflects the received  signal as a passive array. Besides, active relay usually operates in half-duplex mode and  is thus less spectrum efficient than IRS that operates  in full-duplex mode. Although full-duplex relay is also implementable, it requires advanced strong  self-interference cancellation techniques  that are costly to implement.
Second, different from the  traditional backscatter communication such as the  radio frequency identification (RFID)  tag that communicates with the reader  by modulating its reflected  signal  sent from the reader, IRS is used  to facilitate  the existing communication link instead of sending  any information of its own. As such,  the reader  in backscatter communication needs to implement self-interference cancellation at its receiver to decode the tag's message  \cite{griffin2009complete}.  By contrast, in IRS-aided  communication, both the direct-path and reflect-path signals may carry the same useful information and thus can be   coherently added at the receiver to improve the signal strength for decoding. Third, IRS is also different from the active surface based massive MIMO \cite{hu2017beyond} due to their different array architectures (passive versus active) and operating mechanisms (reflect versus transmit). 

{Despite its many  benefits, the IRS-aided  wireless network constitutes both active (BS, AP, user terminal) and passive  (IRS) components, thus differing  significantly from the traditional network comprising  active components only. This thus motivates this article to provide an overview on IRS, including its  signal model, hardware architecture, passive  beamforming design, channel acquisition, node deployment, and so on.  In particular, the  main challenges and their potential solutions for designing and implementing IRS-aided wireless networks  are  highlighted to inspire  future research. Numerical results are also provided to validate the effectiveness of IRS in representative wireless applications. }
\vspace{-0.3cm}
\begin{figure*}[!t]
\centering
\includegraphics[width=0.85\textwidth]{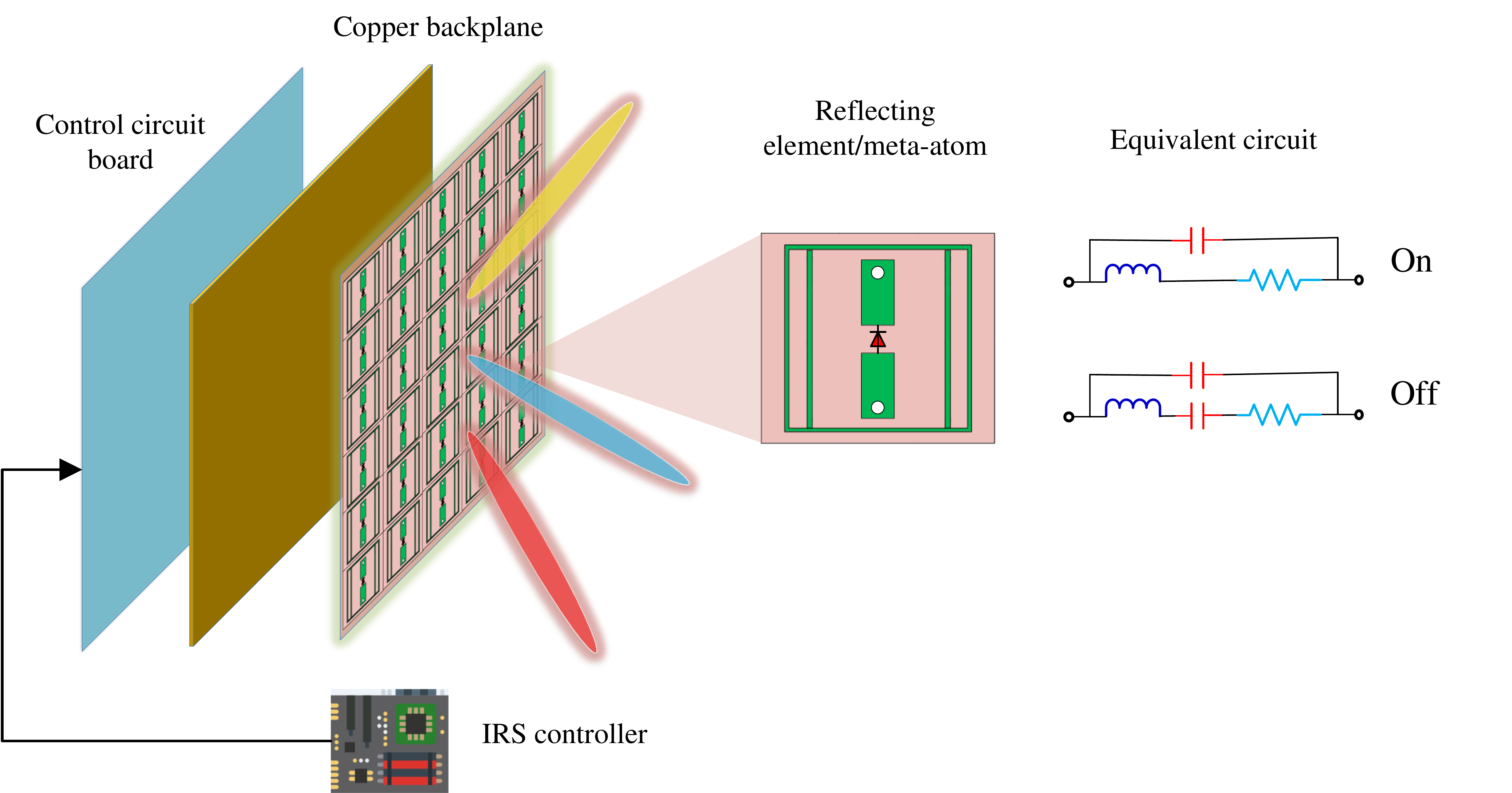}
\caption{Architecture of IRS. } \vspace{-0.6cm}\label{system:hardware}
\end{figure*}
\section{Signal Model and Hardware Architecture}

In this section, we first provide the general signal  model for IRS's reflection,  and then discuss the IRS's  hardware implementation and resultant  constraints on the design of IRS's reflection coefficients in practice.
\vspace{-0.3cm}

\subsection{Signal Model}

  As shown in Fig. \ref{SecI:APP} (a), the composite channel from the BS to the user through each element of  the IRS is a concatenation of three components, namely BS-IRS link, IRS's reflection, and IRS-user link. Such a composite channel is usually referred to as dyadic backscatter channel  in RFID communications \cite{griffin2009complete}, which  behaves different from the conventional point-to-point direct  channel. Specifically, it resembles a keyhole/pinhole propagation where each element of the IRS  receives the superposed multi-path signals from the transmitter, and then scatters the combined signal with adjustable  amplitude and/or  phase as if from a single point source, thus leading to a ``multiplicative'' channel model.

Mathematically,  the reflected signal by  the $n$th element of the IRS, denoted by $y_n$, is given by multiplying the corresponding incident  signal, denoted by $x_n$,   by a complex coefficient, i.e.,  $y_n = \beta_n e^{j\theta_n}x_n,  n=1,\cdots, N,$ where  $\beta_n \in [0, 1]$ and $\theta_n\in [0, 2\pi)$ specify  the  reflection coefficient and control the reflected signal's amplitude (or attenuation due to passive reflection)  and phase shift, respectively, and $N$ denotes the total number of elements of the IRS.
  By smartly adjusting  the reflection  coefficients, the IRS can spatially control the reflected signal to achieve different purposes.
  For example, to maximize the received power of the user in a dead zone  in Fig. \ref{SecI:APP} (a),  all elements of the IRS  should set their reflection amplitude to the maximum value of one (i.e., $\beta_n=1, \forall n$) for maximum signal reflection; whereas to achieve signal/interference cancelation in  Fig. \ref{SecI:APP} (b) or (c), the reflection amplitude of the elements may not necessarily be equal to the maximum value, and can be set different over the  elements.

\vspace{-0.3cm}

\subsection{Hardware Architecture}

The hardware implementation of IRS is based on the concept of  ``metasurface'', which is made of  two-dimensional (2D) metamaterial that is digitally controllable  \cite{cui2014coding}. Specifically, the metasurface is a planar array consisting of a large number of elements or so-called meta-atoms with  electrical thickness in the order of the  subwavelength of the operating frequency of interest  \cite{liaskos18cm}.
By properly designing the elements, including geometry shape (e.g., square or split-ring), size/dimension, orientation, arrangement, etc.,  their individual signal response (reflection amplitude and phase shift) can be modified  accordingly.
In wireless communication applications, the reflection coefficient of each element should be tunable to cater for dynamic wireless channels arising from the user mobility, thus requiring reconfigurability in real time. This can be achieved  by leveraging electronic devices such as 	positive-intrinsic-negative (PIN) diodes, field-effect transistors (FETs), or micro-electromechanical system (MEMS) switches.

As shown in Fig. \ref{system:hardware}, a typical architecture of IRS may consist of three layers and a smart controller. In the outer  layer,  a large number of metallic patches (elements)  are printed on a dielectric substrate to directly interact with  incident signals.  Behind this  layer,  a copper plate is used  to avoid the signal energy leakage. Lastly, the inner layer is a control  circuit board that is responsible for adjusting the reflection amplitude/phase shift of each  element,  triggered by a smart controller attached to the IRS. {In practice, field-programmable gate array (FPGA) can be implemented as the controller,   which also acts as a gateway to communicate and coordinate  with other network components (e.g., BSs, APs,  and user terminals) through separate wireless  links for low-rate information exchange with them.}

One example of an individual  element's structure is also shown  in Fig. \ref{system:hardware},  where a PIN  diode is embedded  in each element. By controlling its biasing voltage via a direct-current (DC) feeding line, the PIN diode can be switched between  ``On'' and ``Off'' states as shown in the  equivalent circuits, thereby generating  a phase-shift difference of $\pi$ in rad \cite{cui2014coding}. 
 As such, different phase shifts of IRS's elements can be realized independently via setting the corresponding  biasing voltages  by the smart controller.  On the other hand, to effectively control the reflection amplitude, variable resistor load can be applied in the element design \cite{yang2017design}. For example, by changing the values of resistors in each element, different  portions of the incident signal's energy are dissipated, thus achieving controllable reflection amplitude in $[0, 1]$.
 In practice, it is desirable to have independent control of the amplitude and phase shift at each element, for which the above circuits  need to be efficiently integrated \cite{yang2017design}. 

\vspace{-0.3cm}
\subsection{Discrete Amplitude and Phase-shift Model}
While continuously tuning the reflection amplitude and phase shift of each IRS's element is certainly advantageous for communication applications, it is  costly to implement in practice because  manufacturing such high-precision elements requires sophisticated design and expensive hardware, which may not be a scalable solution as the number of elements becomes very large. For example, to enable $16$ levels of phase shift as shown in Fig. \ref{system:hardware},   $\log_216=4$ PIN diodes need to be integrated to each element. This not only makes the element design very challenging due to  the limited element size, but also requires more controlling pins from the IRS controller to excite the large number of PIN diodes.
As such,  for practical IRSs that usually have a large number of elements, it is more cost-effective  to implement only discrete amplitude/phase-shift levels requiring  a small number of control bits for each element, e.g., $1$-bit for two-level (reflecting or absorbing) amplitude control, and/or two-level ($0$ or $\pi$) phase-shift control \cite{cui2014coding,wu2018IRS_discrete}.
Note that such coarsely quantized amplitude/phase-shift  design inevitably  causes misalignment of IRS-reflected and non-IRS-reflected  signals at designated  receivers and thus results in certain  performance degradation. 

\vspace{-0.2cm}
\section{Main Design Challenges}
Besides the hardware aspect,  we present in this section other main challenges in designing and implementing IRS-aided wireless networks from the signal processing and communication  perspective, including  passive  beamforming design, IRS channel acquisition, and IRS deployment.
\vspace{-0.3cm}
\subsection{Passive Beamforming Design}  

One challenge of designing the passive beamforming by IRS in practice  lies in the  aforementioned discrete amplitude and phase-shift levels of each element. They  result in exponentially growing complexity orders in terms of the number of IRS elements, $N$, for searching the optimal amplitude/phase-shift discrete values, thus rendering the optimization problem to be NP-hard as $N$ becomes large  \cite{wu2018IRS_discrete}.
As such, a practical  approach is to firstly relax such constraints and  solve the problem with  continuous amplitude/phase-shift values,  then quantize the obtained solutions to their nearest values in  the corresponding discrete sets.  While this approach is generally able to reduce the computation time significantly to polynomial orders of $N$, it may suffer various loss in performance as compared to the continuous-value  solution due to quantization errors, depending on the number of quantization levels as well as $N$, and is also generally suboptimal for the original discrete optimization problem.  To further improve the performance  of the above  approach, the heuristic alternating optimization technique can be applied to iteratively optimize the discrete amplitude/phase-shift values of each element by fixing those of all the others at each iteration  \cite{wu2018IRS_discrete}.

On the other hand, the passive reflect beamforming of IRS in general needs to be jointly designed with the transmit beamforming of other active components in the network such as BSs so as to optimize the network performance. For instance, in Fig. \ref{SecI:APP} (a) where  the BS-user direct link is severally blocked, the transmit beamforming of the BS  ought to point towards  the IRS to maximize its signal reflection for serving  the user. In contrast, when the signal attenuation of the BS-user link is comparable to that of the IRS-reflected  link,   the transmit beamforming  of the  BS should be properly designed to strike a balance between the user's and IRS's directions.  In the above cases, the reflection amplitude of all elements of the IRS  should be set to the maximum value to achieve maximum signal reflection, while the phase shifts need to be tuned based on all channels such that the reflected signal by the IRS can be added constructively at the user's receiver with the direct (non-IRS-reflected) signal from the BS.

For the more general multiuser setup, an IRS-aided  system benefits  from not only the beamforming  of the desired signal  but also the suppression of multiuser interference. For example,  the user closer to the IRS  in Fig. \ref{SecI:APP} (c) can  tolerate more interference from a neighboring  BS,  because  the IRS's reflect beamforming (via both amplitude and phase-shift control)  can be designed  such that the interference reflected by the IRS  can add destructively with that directly  from the interfering BS  to maximally cancel it at the user's receiver. This in turn provides more flexibility for designing the transmit beamforming at the neighboring  BS for  serving  other users outside the IRS's covered region. Despite the above benefits, the active and passive beamforming designs are in general closely coupled and their joint design usually leads to complicated optimization problems that are hard to be solved optimally and efficiently. To reduce such high complexity, alternating optimization can be applied to obtain suboptimal  solutions, by iteratively optimizing one of the transmit and reflect beamforming with the other being fixed, until the convergence is reached \cite{JR:wu2018IRS}.  Furthermore, wireless networks generally operate in wideband channels with frequency selectivity. While active BSs can use digital processing  in frequency domain such as digital beamforming or hybrid digital/analog beamforming to deal with the frequency-selective channel variation \cite{lu2014overview}, it is practically difficult to implement such advanced signal processing for the passive IRS. As a result,  the reflection coefficients of IRS need to balance the  channels at different frequency sub-bands, which further complicates the joint active and passive beamforming optimization.

 Some interesting results have been reported  in this new  direction recently \cite{wu2018IRS,JR:wu2018IRS,huangachievable,wu2018IRS_discrete}.  Prior works \cite{wu2018IRS,JR:wu2018IRS} revealed that in an IRS-aided single-user system, the received power of the user increases asymptotically in the order of  $\mathcal{O}(N^2)$ as the number of reflecting elements $N\rightarrow \infty$. In other words, every  doubling of $N$ achieves about 6 dB power gain in the large-$N$ regime.  The fundamental reason behind such a ``squared law'' of $N$ is that the IRS not only achieves a power  gain of $\mathcal{O}(N)$ by reflect beamforming (similarly like the transmit beamforming with $N$ active antennas in  massive MIMO \cite{lu2014overview}), but also captures another power  gain of $\mathcal{O}(N)$ due to its large aperture for collecting the received signal energy from the BS (which is not available in massive MIMO).    Although such an appealing  power scaling law of IRS can also be obtained theoretically by active MIMO relay with a large number of ($N$) transmit/receive antennas, their full-duplex operation is required which is practically difficult to implement due to the large-dimension MIMO self-interference cancellation.
Moreover, compared to the ideal case with continuous phase shifts, it was shown in \cite{wu2018IRS_discrete} that by using IRS with $b$-bit controlled (or $2^b$-level) uniformly quantized phase shifts, the same asymptotic power scaling law of $\mathcal{O}(N^2)$ can be achieved, while only a constant power loss as a function of $b$ is incurred,  which becomes insignificant  as compared to $\mathcal{O}(N^2)$ and thus can be ignored as $N \rightarrow \infty$.

\subsection{IRS Channel Acquisition}
The various performance gains brought by the passive beamforming of IRS in general require the accurate knowledge of the channels between the IRS and the involved BSs and users. Note that by  turning the IRS into the absorbing mode (i.e., $\beta_n=0, \forall n$), the channel state information (CSI)  of BS-user links without the IRS  can be obtained by applying the conventional  channel estimation methods \cite{lu2014overview}.
Depending on whether receive RF chains are implemented for the elements of IRS  or not, the acquisition of CSI between the IRS and BSs/users  can be  classified into the following two categories.

{First, although transmit RF chains are removed from the IRS for cost reduction and energy saving, each of its elements can be  equipped with a low-power receive RF chain to enable the sensing capability for channel estimation.} As such, the channels from the BSs/users to IRS can be estimated at the IRS based on their training signals. Furthermore, if the time-division duplexing (TDD) is assumed for the uplink and downlink communications in the network, by leveraging  the channel reciprocity, the channels in the reverse directions from the IRS to BSs/users can also be obtained.  To reduce the number of receive RF chains at the IRS, the  \emph{sub-array} technique can be applied where each sub-array consists of a cluster of neighbouring  elements  arranged vertically and/or horizontally and each cluster is equipped with one receive RF chain for channel estimation. Accordingly, the reflection coefficients of all elements in each sub-array can be set  to be either the same or different by using proper  interpolation over adjacent sub-arrays.

On the other hand, when receive RF chains are not installed at the IRS, it is infeasible for the IRS to estimate the channels with involved BSs/users.
However, a viable approach for this challenging case may be, instead of estimating the IRS-BS/user channels explicitly, designing the reflection coefficients for IRS's passive beamforming directly based on the feedback from the BSs/users pertaining their received signals that are reflected by the IRS. For example, the codebook-based passive beamforming  can be implemented  where the IRS quickly sweeps its reflect beamforming coefficients  in a pre-designed codebook and  the best beam is then selected based on the BS/user received training signal power. To reduce the complexity and time overhead of real-time training, historical data can be exploited. For example, for the indoor  mmWave  communication, due to the channel sparsity \cite{lu2014overview},  the IRS-BS/user channels are highly correlated in space and thus the optimal IRS beamfoming coefficients for users in nearby locations with the same associated BS are  similar and vary spatially like a smooth function.  To exploit this,  each IRS can maintain a database that records the optimal beams at different user locations in the past or their differentiable channel fingerprints with the BS. Then, to serve a new user whose location or channel with the BS is available, the IRS can leverage its database to efficiently find an initial set of reflect beamforming coefficients by using e.g., interpolation or machine learning based methods. Such beamforming  coefficients can be further  refined in real time such that the signals from the IRS and BS can add more coherently at the user receiver. In this direction,  \cite{taha2019enabling} recently proposed a  machine learning based phase shift design with low training overhead.  

\vspace{-0.3cm}
\subsection{IRS Deployment}

\begin{figure}[!t]
\centering
\includegraphics[width=0.5\textwidth]{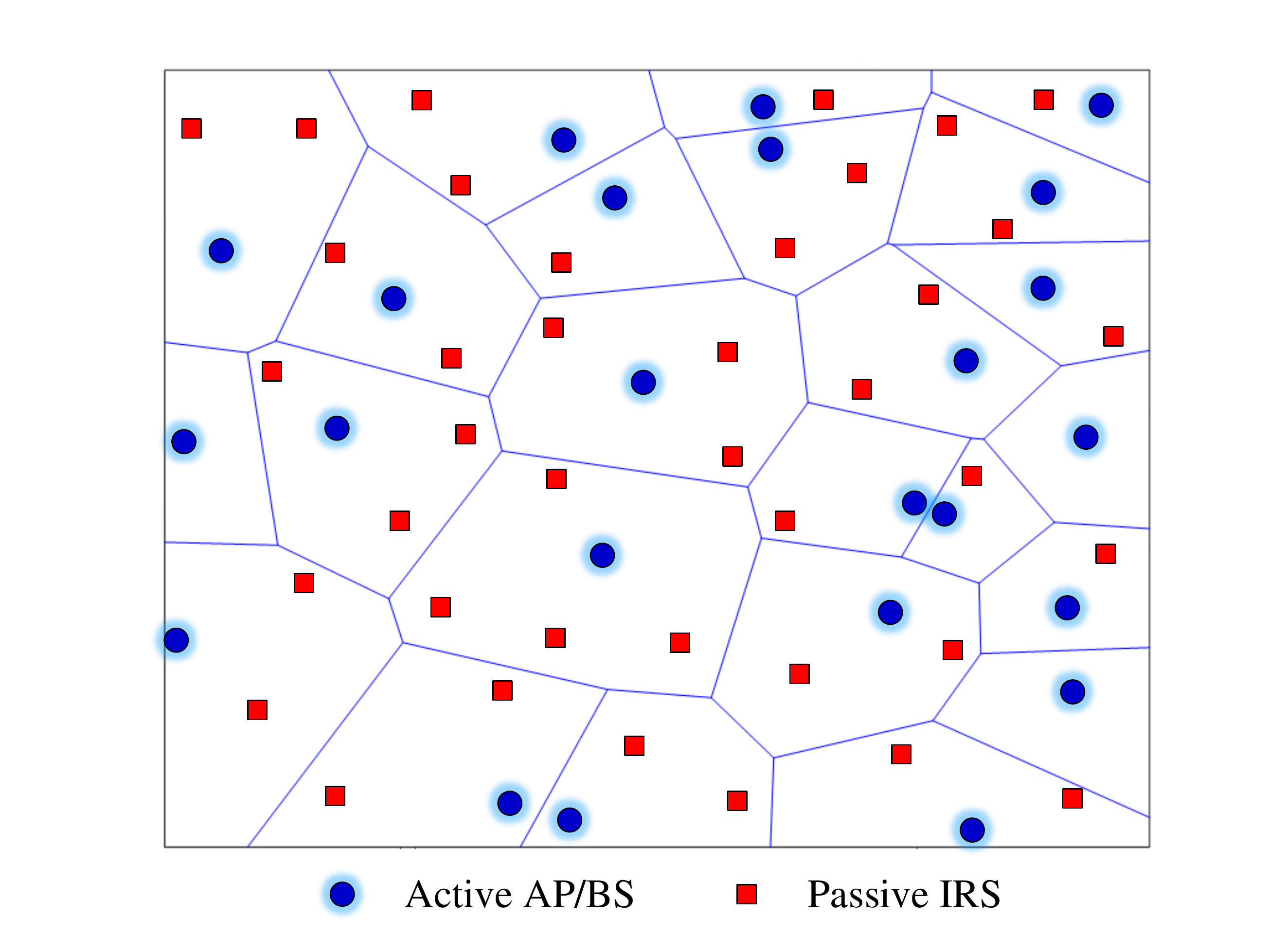}
\caption{Illustration of a hybrid wireless  network with active BSs and passive IRSs. } \vspace{-0.6cm}\label{secIIID:geometry}
\end{figure}

How to judiciously deploy IRSs in a  hybrid  wireless network comprising both active BSs and passive IRSs (as shown in Fig.  \ref{secIIID:geometry})  to optimize its performance is another crucial problem to solve. Generally speaking, this problem should have different considerations as compared to that of deploying active BSs/relays in the traditional wireless network.
{As IRSs are deployed for local coverage only, their operating ranges are usually much shorter than those of active BSs/relays, which thus makes it easier to practically deploy IRSs  without interfering each other.} In the following, we provide more detailed discussions on this issue.

First, from the viewpoint of optimizing the performance in a single-cell setup,  the IRS should be intuitively deployed at a location with clear LoS from the BS in order to maximize its received signal power for passive beamforming. However, when the IRS needs to support simultaneous transmissions between the BS and the users in its coverage region,  such a straightforward deployment strategy may not work well. This is because  a single LoS path between the IRS and the BS results in low-rank MIMO channel between them and thus very limited spatial multiplexing gain of the overall IRS-aided links between the BS and  served users  \cite{JR:wu2018IRS}. Therefore, the deployed location for IRS is practically preferable to possess a strong LoS path with the BS as well as a sufficiently large number of non-LoS paths for enabling high-rank MIMO channel, so as to resolve the above trade-off.   {Besides,  the deployment of IRS should also take into consideration the spatial user density, i.e., with a high priority to be deployed in  hot-spot areas with large number of users, as well as the inter-cell interference issue, e.g., when there is an urgent  need to deploy an IRS  near the boundary of two adjacent cells to help cancel the co-channel interference between them as shown in Fig. \ref{SecI:APP} (c).}

In practice, the propagation  environment may be complicated  and  each IRS can be associated with multiple  BSs. In such scenarios, using good heuristics alone for deploying IRS may be ineffective, while an exhaustive search for the optimal location requires the global CSI at all locations, which is practically difficult to obtain. Ray-tracing based methods can be used to estimate such CSI, but they are computationally costly and also require site-specific information  (such as  building/floor  layout for indoor communication). As such, how to achieve autonomous deployment of IRSs to predict the most suitable locations for them is a new problem of high practical interest.  One promising approach to solve this problem is by leveraging the machine learning techniques, such as deep learning (DL). {For example, in the training phase, we can deploy IRSs at some properly selected reference locations and collect key performance indicators such as received signal strength measured at different user locations. Such IRS locations and corresponding performance indicators are then used to train a DL-based neural network as the output and input, respectively. Next, in the deployment phase, with the desired performance indicators as the input, the trained DL network is used to predict a set of locations for deploying IRSs. After deploying IRSs at these locations, a new set of performance indicators can be collected and used to further train the DL network to improve its prediction accuracy in the future.   }

Last, from the perspective of designing a multi-cell wireless network, the following interesting new problem arises: what is the fundamental capacity  limit of  IRS-aided wireless networks? Although UDN is undoubtedly the most significant driver for the astounding capacity increase in 5G networks, it is known that for wireless networks with active BSs only, increasing the spatial  density of BSs  beyond a certain threshold will reduce the network capacity asymptotically  to zero, due to the  increasingly more severe  interference \cite{andrews2016we}. However, this pessimistic result  is not applicable to the IRS-aided wireless network with hybrid active BSs and passive IRSs, as shown in Fig. \ref{secIIID:geometry}.  Since IRSs are passive and thus do not increase the network interference level, while they help enhance the user performance in their locally covered areas by   intelligent reflection, it is expected that the network capacity will be significantly improved by adding IRSs as compared to the conventional network with active BSs only. It is also plausible that increasing the spatial density of passive IRSs may not lead to the network capacity degradation due to increased interference in the conventional network. This thus opens a new avenue for investigating the optimal wireless network architecture and its new capacity limit in future work.

\vspace{-0.1cm}
\section{Numerical Results and Discussion}
{Numerical results are provided in this section to demonstrate the effectiveness of the proposed IRS in creating  ``signal hotspot'' and ``interference-free zone'' shown in Fig. \ref{SecI:APP}, respectively.}
We consider a BS with $M$ antennas, an IRS with $N$ elements, and
 one single-antenna user, with their locations shown in Fig. \ref{simulation:pow2} (a).    Denote the horizontal distance between the BS and  user  by $d$ meter (m).
It is assumed that the BS-IRS channel is dominated by the LoS link due to proper deployment with the path loss exponent of 2.2, whereas both BS-user and IRS-user channels are assumed to follow Rayleigh fading with path loss exponent of 3.2.  The noise power at the user receiver is $-80$\,dBm.


\vspace{-0.3cm}
\subsection{ Signal Power  Enhancement and Scaling Law} 
\begin{figure*}[t]
\centering
\subfigure[Simulation setup of the single-user case.]{\includegraphics[width=0.495\textwidth]{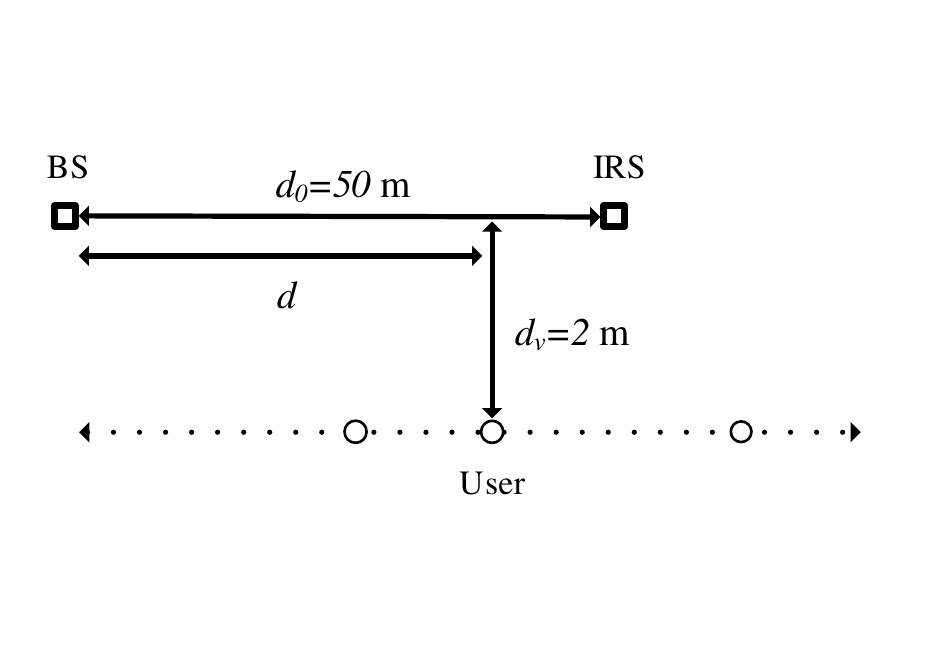}}
\subfigure[BS transmit power versus BS-user horizontal distance.]{\includegraphics[width=0.495\textwidth]{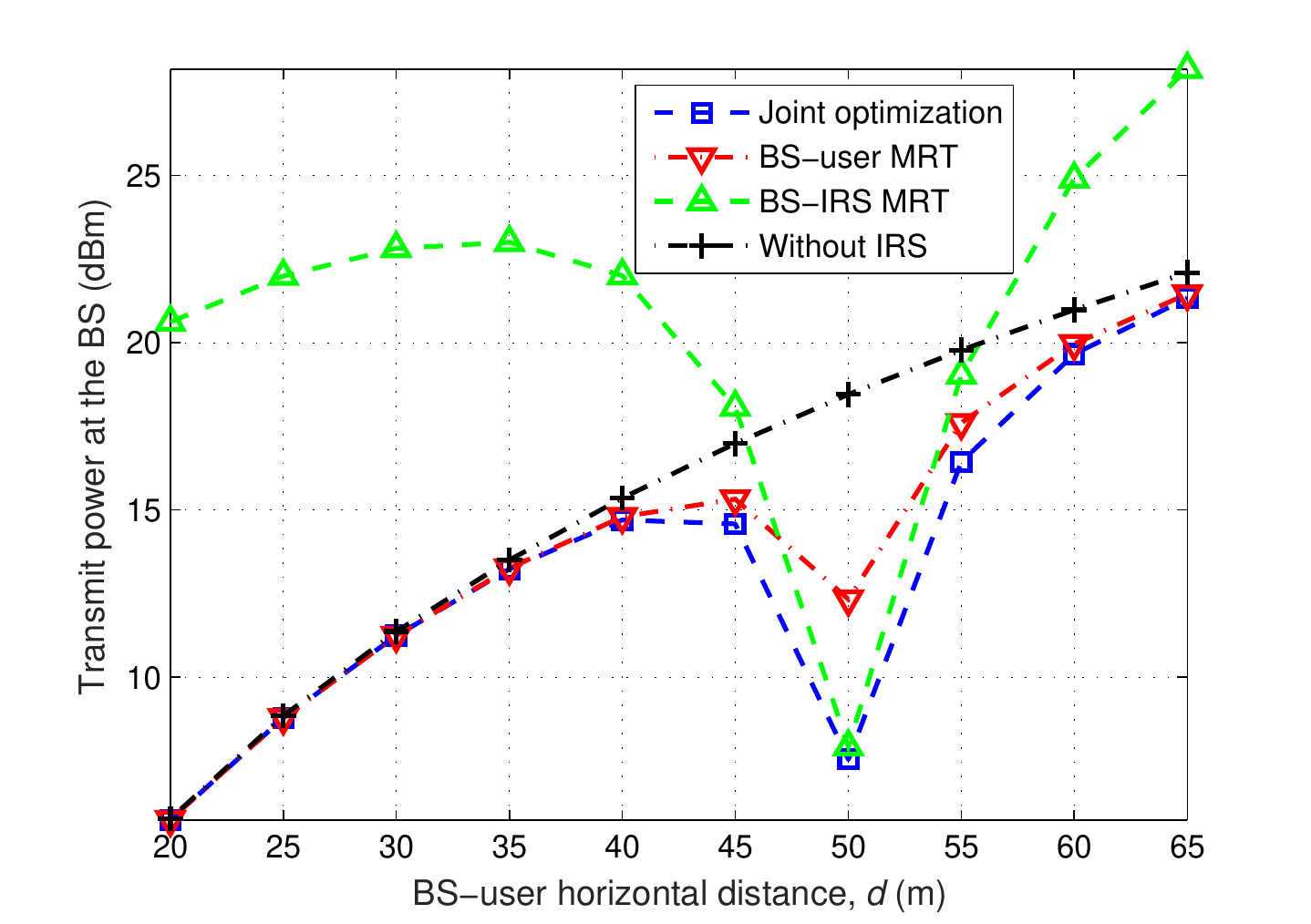}}
\subfigure[BS transmit power versus $N$. ]{\includegraphics[width=0.485\textwidth]{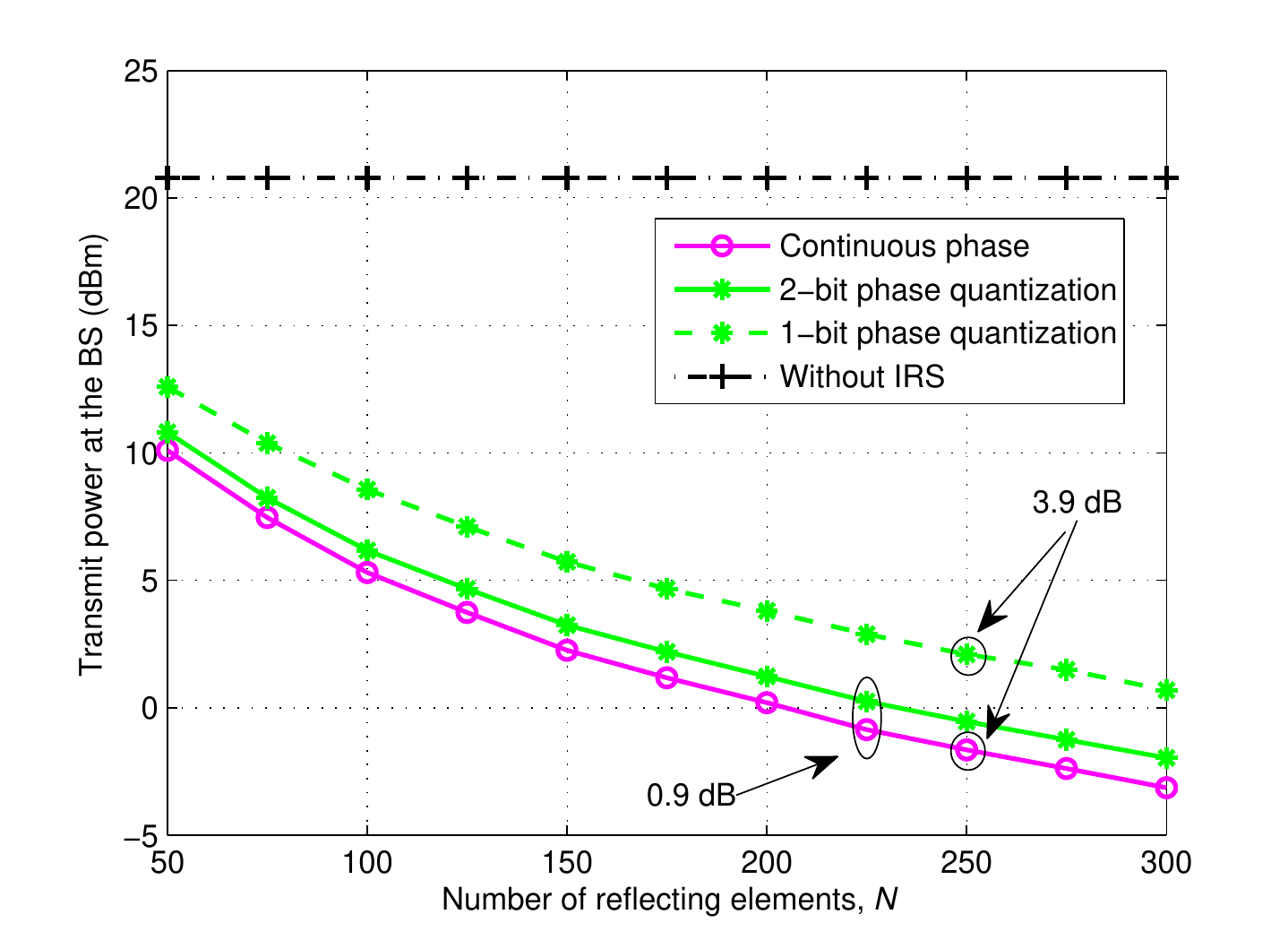}} 
\subfigure[Normalized interference power versus $N$. ]{\includegraphics[width=0.495\textwidth]{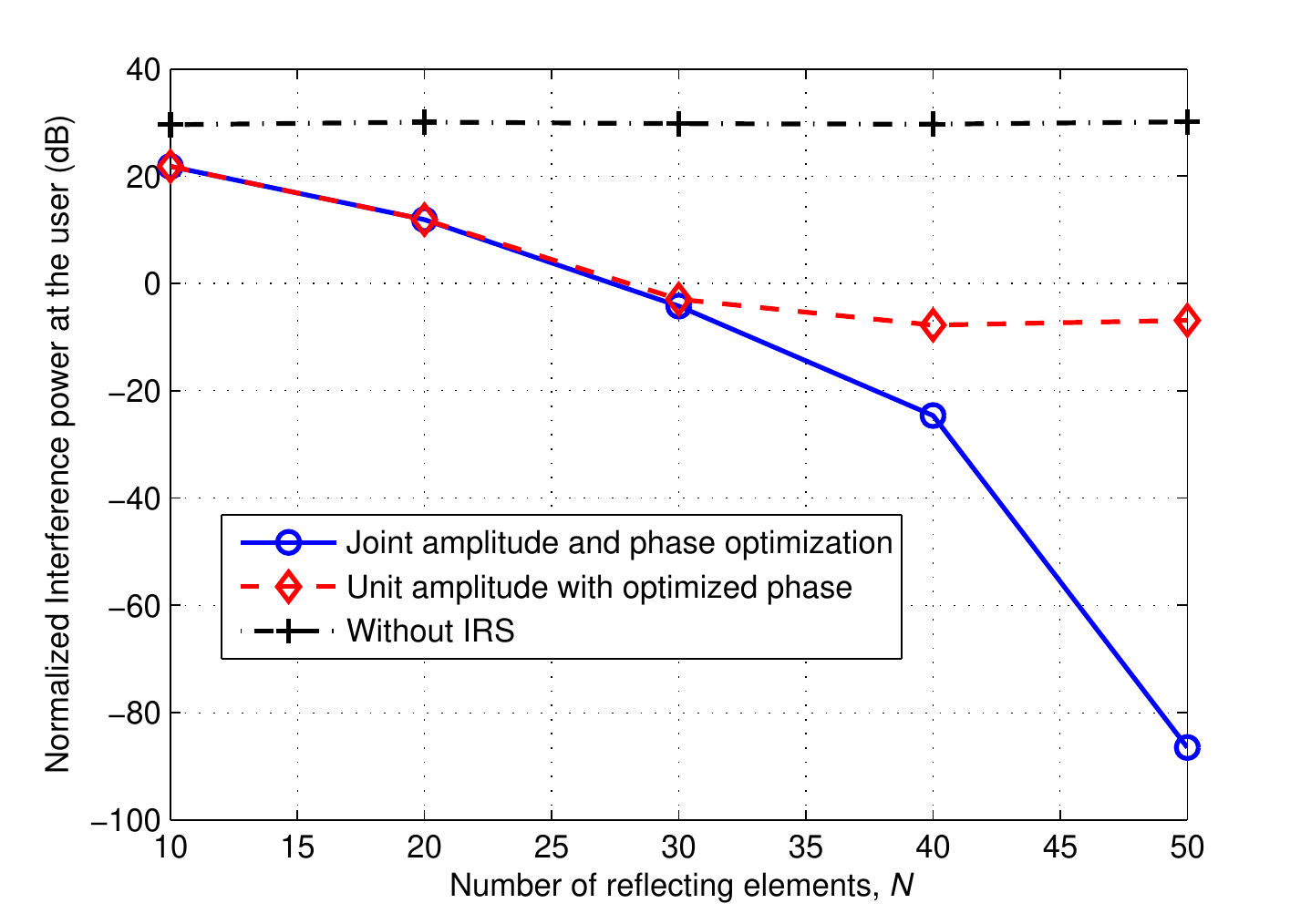}}
\caption{Performance of IRS-aided wireless networks. } \label{simulation:pow2} \vspace{-6mm}
\end{figure*}

{To demonstrate the signal power enhancement capability of  IRS,  we assume that  the user in Fig. \ref{simulation:pow2} (a) needs to be  served by the BS with the IRS's help, similar to the scenario in  Fig. \ref{SecI:APP} (a).} Thus, the signal reflected by the IRS should constructively add with that from the BS-user direct link.   {We compare the following four schemes under the setup of  $M=5$ and $N=40$: 1) Joint optimization where the active BS transmit beamforming and passive IRS reflect beamforming are jointly optimized as in \cite{JR:wu2018IRS}; 2) BS-user maximum-ratio transmission (MRT) where the BS beams towards the BS-user channel; 3) BS-IRS MRT where the BS beams  towards the BS-IRS rank-one channel; and 4)  Benchmark scheme without the IRS where the BS beams  towards the BS-user channel.}
  As shown in  Fig. \ref{simulation:pow2} (b), by varying the value of $d$, we examine the minimum  transmit power required at the BS for achieving a target user signal-to-noise ratio (SNR) of 20 dB.  First, it is observed that for the scheme without IRS,  moving  the user farther away from the BS leads to  higher transmit power due to the increased signal attenuation. However, this problem is alleviated by deploying the IRS, which helps significantly improve the SNR when the user is near to it.
As a result, the user near either the BS (e.g., $d=25$ m)  or IRS (e.g., $d=50$ m) requires  lower transmit power than a user far away from both of them (e.g., $d=40$ m).   This demonstrates the practical usefulness of  IRS in creating a ``signal hotspot'' in its vicinity.
 Furthermore, compared to other heuristic BS transmit beamforming schemes, the joint active and passive beamforming design also achieves substantial power saving at the BS. {Note that the BS-IRS MRT scheme performs poorly when the user is out of the IRS's coverage since in this case the user received signal power is dominated by that of the BS-user direct link and the BS-IRS MRT is thus ineffective as it beams the BS power toward the IRS instead of the user.}

In Fig. \ref{simulation:pow2} (c), we show the performance of the IRS where each of its elements reflects with a constant amplitude (assumed to be the maximum value of one), but a practical  $b$-bit uniformly quantized phase shifter. The  BS transmit power  is plotted versus the number of reflecting elements of the IRS, $N$, with $d=50$ m. 
First,  it is interesting to observe  that for the ideal case of continuous phase ($b=\infty$), the BS transmit power scales down with $N$ approximately in the order of  $\mathcal{O}(N^2)$. For example, for the same user SNR,  a transmit power of 2.5 dBm is required at the BS when $N=150$ while this value is reduced to about $-3.5$ dBm when  $N=300$, which suggests around  6 dB gain by doubling $N$.
{Second, one can observe that the performance loss of using finite-level phase shifters with  $b=1$ bit or $b=2$ bits first increases with $N$ and eventually  approaches a constant value, i.e.,  $3.9$ dB or $0.9$ dB, which are consistent with the results given  in \cite{wu2018IRS_discrete}.}

 \vspace{-0.3cm}
\subsection{Interference Suppression}
{Next, we demonstrate the interference suppression capability  of  IRS, by considering now the BS  in Fig. \ref{simulation:pow2} (a) is a neighbouring transmitter that causes co-channel interference to the user with $d=50$ m, and thus the IRS is deployed to help suppress its received  interference from this BS, for the scenario shown in   Fig. \ref{SecI:APP} (c).
This setup also resembles  the physical layer security scenario  in Fig. \ref{SecI:APP} (b), where the user is an  eavesdropper and its received signal from the legitimate transmitter (BS) needs to be canceled with the help of the IRS.}
For simplicity, we assume that $M=1$ and the transmit power of the BS is 30 dBm. For comparison,   we plot in Fig.  \ref{simulation:pow2} (d) the interference power (normalized by the noise power) at the user versus  $N$ for three schemes,
i.e., the scheme with jointly optimized continuous amplitude and phase shifts, the scheme with unit amplitude  and optimized continuous phase shifts, and the scheme without IRS.  {The first two schemes can be efficiently realized  by applying the  semidefinite relaxation (SDR) technique.} It is first observed that as compared to the scheme without IRS, the interference power is substantially reduced even by adjusting the phase shifts of IRS's elements  solely. Moreover, with jointly optimized continuous amplitude and phase shifts, it is observed that the co-channel interference can be more effectively canceled as compared to the case with fixed amplitude, especially when $N$ is sufficiently large.  This is because with the additional amplitude control, the IRS is able to impose  an opposite interference signal at the user to perfectly cancel  that  from the BS-user link, thus creating a virtually  ``interference-free  zone''.  


\section{Conclusions}
{In this article, we provide an overview of the promising IRS technology for achieving smart and reconfigurable environment in future wireless networks. In particular, the  IRS can sense  the wireless environment and accordingly adjust its reflection coefficients dynamically  to achieve different functions by leveraging advanced signal processing and  machine learning algorithms.}
As IRS-aided wireless networks are new and remain largely unexplored,  it is hoped that this article would provide a useful and effective guidance for the future work on investigating them in various aspects.  In particular, we foresee that the integration of IRSs into future wireless networks will fundamentally change their architecture from the traditional  one with active  components solely to a new hybrid one with both active and passive components co-working in an intelligent way, thus opening  fertile directions for future research.

%
%
%

\bibliographystyle{IEEEtran}
\bibliography{IEEEabrv,mybib3}
\end{document}